\documentclass{ws-procs9x6}

\begin{document}

\title{RECENT RESULTS OF THE OPERA EXPERIMENT}

\author{F. PUPILLI\footnote{on behalf of the OPERA Collaboration}$^+$}

\address{INFN-Laboratori Nazionali del Gran Sasso,\\
I-67100 Assergi (L'Aquila), Italy\\
$^+$E-mail: fabio.pupilli@aquila.infn.it}

\begin{abstract}
The OPERA experiment aims at the direct confirmation of the leading oscillation mechanism in the atmospheric sector looking for the
appea\-ran\-ce of $\nu_{\tau}$ in an almost pure $\nu_{\mu}$ beam (the CERN CNGS beam). In five years of physics run the experiment 
collected $17.97 \times 10^{19}$ p.o.t. The detection of $\tau$s produced in $\nu_{\tau}$ CC interactions and of their decays is 
accomplished exploiting the high spatial resolution of nuclear emulsions. Furthermore OPERA has good capabilities in detecting 
electron neutrino interactions, setting limits on the $\nu_{\mu} \rightarrow \nu_{e}$ oscillation channel. In this talk the status
of the ana\-ly\-sis will be presented together with updated results on both oscillation channels.
\end{abstract}


\bodymatter

\section{The OPERA detector}
The OPERA (Oscillation Project with Emulsion-tRacking Apparatus) experiment\cite{proposal} has been conceived for the confirmation of
the $\nu_{\mu} \rightarrow \nu_{\tau}$ oscillation in appearance mode, through the direct detection of the short-lived $\tau$ lepton produced
in charged current (CC) interactions of $\nu_{\tau}$.
It is installed in the underground Gran Sasso National Laboratory and was exposed to an almost pure $\nu_{\mu}$ beam (CNGS\cite{cngs})
produced by the CERN SPS, 730 km far away form the detector.
The OPERA detector\cite{detector2009} is a hybrid apparatus, with nuclear emulsions complemented by electronic detectors. 
The core of the experiment is made of modular units 
called bricks. A brick
is a ``sandwich'' of 56 1 mm thick lead plates, acting as target material, and 57 300 $\mu$m thick nuclear emulsion films, acting
as high precision tracking devices. 
About 150000 bricks, for a total mass of about 1.25 kton, are arranged in vertical structures transverse to the beam direction called
walls. Each wall is followed by two planes of plastic scintillator strips composing the Target Tracker (TT). At the downstream face of the
brick, two additional emulsion films (Changeable Sheets, CS) can be removed without disassembling it and act as interface between the
electronic detectors and the emulsions.
The target section is followed by a muon spectrometer, composed of a dipole magnet instrumented with Drift Tubes and Resistive Plate
Chambers. This structure is replicated in two subsequent Super-Modules.

\section{Event reconstruction and analysis}
In five years of CNGS operations, from 2008 to 2012, OPERA collected $17.97 \times 10^{19}$ protons on target (p.o.t.).

Electronic detector data are processed in order to reconstruct events on-time with the CNGS beam and an algorithm classifies them
as occurred inside (contained events) or outside (external events) the OPERA target. 
Only contained events are used for oscillation studies.
Furthermore the identification of (at least) a three dimensional track as a muon, by e\-va\-lua\-ting the amount of traversed material, 
allows the separation of the sample into CC-like (also called 1$\mu$) and NC-like (0$\mu$) events. The muon charge and momentum are also
measured, using the magnetic spectrometers\cite{eledet2011}.
For each event, the brick with the highest probability to contain the neutrino interaction is selected by a brick finding algorithm
and the associated CS are analysed. If a signal compatible with TT data and with an upstream neutrino interaction is found, the corresponding brick
is dismounted and the emulsion films are analysed, by means of fast automated optical microscope systems providing sub-micron spatial accuracy.
Tracks found in CS are looked for in the most downstream film of the brick and then followed back from film to film, until
they are not found in three consecutive films; this stopping point is considered as the signature either for a primary or a secondary
vertex. A dedicated scanning of $\sim$ 2 cm$^{3}$ around the stopping point and a vertexing algorithm allow the vertex confirmation and
provide the picture of the neutrino interaction.
This picture is supplemented by a decay search procedure, aimed at the detection of decay topologies and/or secondary interactions,
searching for large kink angle along tracks or for tracks with large impact parameter (IP) with respect to the primary vertex (since
the IP for primary tracks usually does not exceed 10 $\mu$m). When a decay vertex is found, a full kinematical analysis is performed:
the momentum of charged particles is estimated by measuring angular deviations due to the Multiple Coulomb Scattering in lead\cite{MCS} ,
with a resolution of $\sim$ 22\%; exploiting calorimetric techniques, made possible by the 10 X$_{0}$ thickness of the brick,
the energy of electromagnetic showers coming from electrons or $\gamma$-ray conversions can also be measured.
Since charmed particles have lifetime and decay topologies similar to those of a $\tau$ lepton, $\nu_{\mu}$ CC interactions with the
production of a charmed particle constitute the most relevant background for all $\tau$ decay channels, but also an important
benchmark for the $\tau$ finding efficiency. In particular, observed charmed events in the 2008-2010 data (50 events) are
in good agreement with Monte Carlo expectations (53 $\pm$ 5 events).

\section{$\nu_{\mu} \rightarrow \nu_{\tau}$ oscillation search}
In the presently analysed sample, three $\nu_{\tau}$ candidates were observed.

The first one was found in the 2008-2009 0$\mu$ data sample and described in detail in Ref.~\refcite{first_tau}. The event has seven
prongs at the primary vertex; six of them are incompatible with the muon hypothesis by detecting a hadronic
interaction or by muon--range consistency checks. The seventh track, associated to a $\tau$ lepton, displays a kink topology and the daughter
track is identified as a hadron through its interaction. Two $\gamma$-rays have been also found pointing to the decay vertex and their
measured invariant mass ($120\pm20(stat.)\pm35(syst.)$ MeV$/$c$^{2}$) is compatible with the $\pi^{0}$ mass; their combination with
the secondary hadron, assumed to be a $\pi^{-}$, gives an invariant mass of $640^{+125}_{-80}(stat.)^{+100}_{-90}(syst.)$ MeV$/$c$^{2}$.
Therefore the decay mode is compatible with $\tau \rightarrow \rho(770) \nu_{\tau}$.

The second $\nu_{\tau}$ candidate was found in the 2011 0$\mu$ data sample\cite{second_tau} and consists of a two prong primary vertex.
One of the tracks disappears after the first downstream brick and has a momentum of $2.8^{+0.7}_{-0.7}$ GeV$/$c; it was associated
with a hadron, being incompatible with a muon track because of its range. The second track, associated with the $\tau$ lepton,
exhibits a three prong decay topology; the daughter particles were identified
as hadrons on the basis of muon--range correlation or by directly ob\-ser\-ving their hadronic interaction. The kinematical
analysis, together with the topological one, satisfies the criteria for the $\tau \rightarrow 3h$ decay channel.

\begin{figure}
\begin{center}
\psfig{file=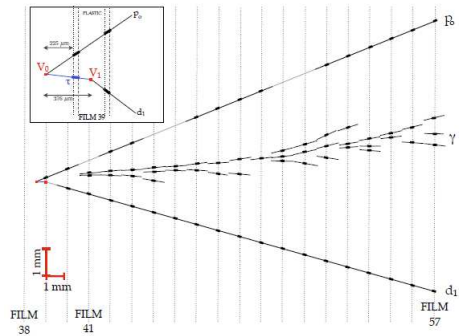,width=2.4in}
\end{center}
\caption{Display of the third $\nu_{\tau}$ candidate event. In the inset is shown a zoomed view of the primary and secondary vertices
region. See the text for a detailed description.}
\label{tau_display}
\end{figure}

The third candidate has been observed in the 2012 1$\mu$ data sample\cite{third_tau} . As shown in Fig.~\ref{tau_display}, the primary vertex ($V_{0}$)
is given by two tracks, and by a $\gamma$-ray with a reconstructed energy of $3.1^{+0.9}_{-0.6}$ GeV. Track $p_{0}$ has been
associated with a hadron, since it has a momentum of $0.90^{+0.18}_{-0.13}$ GeV$/$c and it stops in the downstream brick;
therefore the momentum--range correlation strongly disfavours the muon hypothesis. The other track is identified with the $\tau$ lepton
decaying into a muon; indeed track $d_{1}$, the decay daughter, has been matched with the muon track reconstructed by the electronic
detectors, with a measured momentum of $2.8 \pm 0.2$ GeV$/$c and a negative charge assessed with a 5.6 $\sigma$ significance. The negative
charge of the muon strongly suppresses the hypothesis of a charm event with an undetected primary muon. All the kinematical cuts for the
selection of $\tau \rightarrow \mu$ decays are satisfied.

In the analysed sample 0.027, 0.116, 0.021 and 0.020 background events are expected in the $\tau \rightarrow h$, $\tau \rightarrow 3h$,
$\tau \rightarrow \mu$ and $\tau \rightarrow e$ channels respectively, coming essentially from charmed events with an undetected primary muon,
hadronic re-interactions (for the hadronic decay channels) and large angle muon scattering (for the $\tau \rightarrow \mu$ channel).
Taking into account the different signal-to-noise ratio for each decay channel, the three observed candidates give a 3.4 $\sigma$ 
significance to a non-null observation of $\nu_{\mu} \rightarrow \nu_{\tau}$ oscillations.

\section{$\nu_{\mu} \rightarrow \nu_{e}$ oscillation search}
\begin{figure}
\begin{center}
\psfig{file=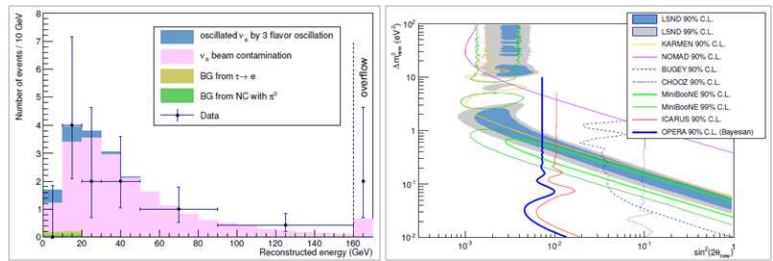,width=4.0in}
\end{center}
\caption{Left panel: reconstructed $\nu_{e}$ energy compared to MC expectations. Right panel: Exclusion plot for the parameters of the
non-standard $\nu_{\mu} \rightarrow \nu_{e}$ oscillation.}
\label{nue}
\end{figure}

A systematic search for $\nu_{e}$ events was applied to 505 0$\mu$ events in the 2008 and 2009 data sample\cite{nue} .
The number of observed $\nu_{e}$ interactions (19 events) is compatible with the expected $\nu_{e}$ from beam contamination and background
(19.8 $\pm$ 2.8). Figure~\ref{nue} (left panel) shows the reconstructed energy distribution of these 
events, compared with the ones
expected from the $\nu_{e}$ beam contamination, the oscillated $\nu_{e}$ from three--flavour oscillation and
the background.
Two different oscillation scenarios are considered. In the standard three--flavour mixing scheme, by introducing a cut at 20 GeV on the
reconstructed energy to improve the signal-to-background ratio, 4.6 events are expected while 4 are observed, and an upper limit
sin$^{2}(2\theta_{13})<0.44$ is derived at the 90\% Confidence Level (C.L.). In the non-standard oscillation framework with a large
$\Delta m^{2}_{new}$ suggested by the LSND and MiniBooNE experiments, by introducing a cut at 30 GeV to improve the sensitivity,
9.4 $\pm$ 1.3 ($syst.$) events are expected from background, while 6 are observed. Given the under-fluctuation of the data, 
a Bayesian approach has been followed to derive the exclusion plot in Fig.~\ref{nue} (right panel) and the upper limit 
sin$^{2}(2\theta_{new})<7.2 \times 10^{-3}$ at 90\% C.L.

\section{Conclusions}
The physics run of OPERA started in 2008 and ended on December 2012, allowing the collection of $17.97 \times 10^{19}$ p.o.t; the
analysis at the emulsion level is still on going. Three $\nu_{\tau}$ candidates have been observed so far and the non-null observation
of $\nu_{\mu} \rightarrow \nu_{\tau}$ oscillations is excluded at 3.4 $\sigma$. The observed number of $\nu_{e}$ interactions is compatible
with the non-oscillation hypothesis, allowing OPERA to set also an upper limit in the parameter space for a non-standard $\nu_{e}$ appearance.

\bibliographystyle{ws-procs9x6}
\bibliography{recent_results_OPERA}

\end{document}